# Methodical Approach for Centralization Evaluation of Modern Automotive E/E Architectures

Lucas Mauser[1], Stefan Wagner[2], Peter Ziegler

[1] Daimler Truck AG, Fasanenweg 10, 70771 Leinfelden-Echterdingen, Germany
`lucas.mauser@daimlertruck.com`
[2] University of Stuttgart, Universitätsstraße 38, 70569 Stuttgart, Germany

**Abstract.** Centralization is considered as a key enabler to master the CPU-intensive features of the modern car. The development and architecture change towards the next generation car is influenced by ADAS, connectivity, infotainment and the consequential need for cyber-security. There is already a high number of papers describing future centralized E/E architectures and technical instruments for centralization. What is missing is a methodical approach to analyze an existing system and find its potential for centralization on the function level. This paper introduces an approach, which serves a system designer or engineer to abstract functions and thus enables to shape a more centralized system architecture. The commonly known E/E architecture designs and the named instruments of current research are the basis for this abstraction. Based on the approach, new system architecture proposals can be set up to discuss and outweigh advantages and disadvantages of those. The approach is validated by applying it step by step to the inlet's derating function of a modern electric vehicle. A following discussion points out that many different factors affect the potential for centralization and centralization may not be the future of every function and system in general.

## 1 Introduction

The domain-oriented E/E architecture is commonly known as state of the practice in established automotive manufacturers such as Daimler, Ford or Renault. Their vertical architectures are characterized by approximately five domain areas each containing a high number of specialized Electronic Control Units (ECUs) resulting in highly distributed functionalities [1]. Those vertical architectures miss flexibility and scalability to satisfy the modern trends in automotive where new players enter the market starting from the scratch and gaining design freedom [2]. Not only Advanced Driver Assistance Systems (ADAS) with their high computing power demand require a redesign of common E/E architectures. Also the connectivity of the modern car with its linkage to services of the world wide web, the TCP/IP protocols and communication bandwidth hungry infotainment systems are pushing the trend of centralization forward [3, 4]. Last but not least, the connectivity of the vehicle involves cyber-security requirements.

Under the high cost pressure of the automotive industry, decentral and well-established platforms are limited in their scalability and are no more able to handle the complexity. The comparison of a Tesla model Y, a VW ID.4 and a Ford Mach E regarding their number of ECUs and communication networks in Table 1 accentuates this statement where Volkswagen and Ford are still captured in their legacy platforms. Instead of a revolution, the established automotive OEMs are in an evolution.

Many Tier-1 suppliers are already serving the need for centralized High-Performance Computing (HPC) ECUs as masters of zone-oriented E/E architectures surrounded by their smart actuators and sensors. This principle separates computing power and I/O to centralize and bundle complex functions into one ECU making complexity manageable [5]. Another technical approach that the Tier-1 suppliers' solutions have in common is the service-oriented design and communication which allows to process the high amount of data that will further increase in the future [5, 6, 7].

The current trend of vertical integration within companies [8] supports the centralization of E/E architectures as know-how is increasing in the staff. In this elaboration, we point out technical and structural approaches to centralize E/E architectures based on related work. In a next step, those approaches are brought together into a methodical approach. Additionally we discuss the limits of centralization critically. Before we deep-dive in approaches for centralization, the characteristics of domain- and zone-oriented E/E architectures are worked out to better classify those approaches.

**Table 1.** Comparison Tesla Model Y, VW ID.4 and Ford Mach E [9]

|  | ID.4 | Model Y | Mach E |
|---|---|---|---|
| ECUs | 52 | 26 | 51 |
| CAN | 7 | 10 | 8 |
| CAN-FD | 6 | Some CAN buses FD capable | 1 |
| Ethernet | 12 | 2 | 4 |
| LIN | 9 masters, 43 slaves | 5 masters, 24 slaves | 13 masters, 44 slaves |
| LVDS | 3 | 10 | 3 |
| Other | - | A2B, BroadR | A2B |

## 2 Dominating E/E Architectures and Trends

### 2.1 Domain-Oriented E/E Architectures

On top of a domain-oriented E/E architecture is the so called domain controller which is a powerful master CPU responsible for controlling and monitoring a dedicated domain, for example, the powertrain domain. The domain controller bundles and consolidates functionality and thus meets the need of increasing complexity and increasing demands of computing power. Below the domain controller, specialized ECUs with less computing power than the domain controller handle function specific tasks which might require a certain hardware or a dedicated location. Those specialized ECUs are connected to its domain controller by typical communication networks like e.g. LIN for simple I/O actuator control and sensor monitoring but also CAN for highly interactive ECUs. To enable higher data rates, CAN-FD with up to 5 Mbit/s is commonly used which might be sufficient for domain-oriented E/E architectures with regards to the actual, centralization-driven development. Automotive Ethernet can enable even higher data rates but it is more seen as backbone communication path between two or more domains [1, 10]. Fig. 1 shows an example of such a domain-oriented E/E architecture. Commonly there are four to five domains in a vehicle which got established. For example, at Mercedes-Benz these are Infotainment and Telematics, the Body and Comfort, ADAS and Powertrain. They represent also the main domains in literature. [1, 10]

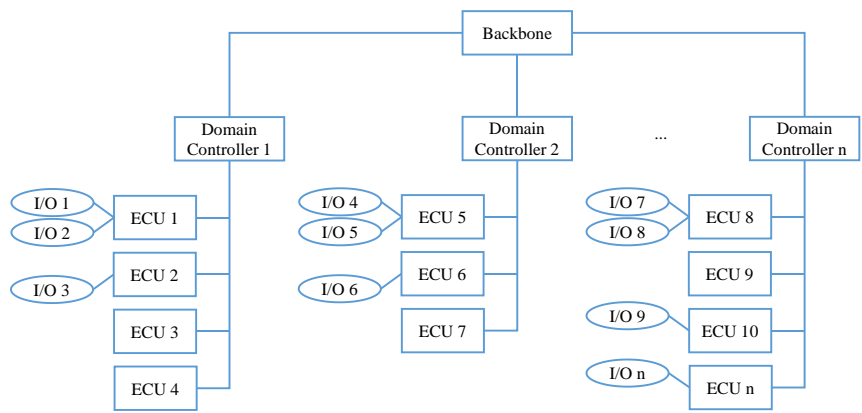

**Fig. 1.** Domain-oriented E/E architecture

### 2.2 Limitations and Challenges

The two domains ADAS and powertrain underline the limitation that a domain-oriented E/E architecture can have. Both domains need to work together closely to enable autonomous driving. There is many information to be exchanged over domain borders by which the functionality suffers from a certain distribution and decentralization [11]. With increasing implementation complexity and number of interfaces, manageability and scalability decrease. At that moment an evaluation to redesign the E/E architecture in the direction of a cross-domain- or zone-oriented E/E architecture will be useful. This kind of architecture and its characteristics will be presented in the following.

### 2.3 Cross-Domain- and Zone-Oriented E/E Architectures

Cross-domain-oriented E/E architectures are one intermediate step towards zone-oriented E/E architectures. The cross-domain controller gets more powerful by accommodating functions of different domains which for example have a lot of common interfaces or common requirements for specialized software and hardware. By combining such synergies in one, a cross-domain controller can increase efficiency in implementation, performance, communication and further serving a better manageability and scalability of vehicle functions [1].

Centralizing one step further brings us to the zone-oriented E/E architecture whose development a lot of Tier-1 suppliers are already pushing forward as stated in the introduction. The zone-oriented E/E architecture is characterized by a central master as a powerful vehicle computer. It combines all the functionalities of the former domain controllers by technical solutions as virtualization, containerization and further which will be investigated later in this elaboration [1, 11].

The rest of the vehicle is physically divided into zones that consist each of a zone controller as gateway to I/O hardware like for example actuators or sensors that cannot be centralized. The gateway zone controllers translate signals for and from the central computer. Direct signal lines and communication channel types must be chosen dependent on the hardware and the required bandwidth [1]. A typical zone-oriented E/E architectures is illustrated in Fig. 2.

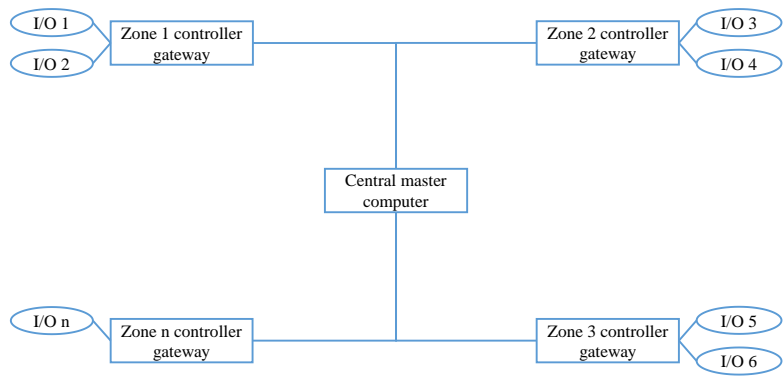

**Fig. 2.** Zone-oriented E/E architecture

### 2.4 Limitations and Challenges

While investigating advantages and disadvantages of centralization in literature, there were by the majority positive aspects of a centralization with focus on the current and future development of the vehicle's technologies. Two disadvantages are emphasized which need to be addressed by technical, methodical and design-specific solutions.

First, a centralized E/E architecture is characterized by a much higher number of wire length and cut leads which can be addressed by a smart design and distribution of the zones. Second, there is the more critical disadvantage of single point of failure which a centralized system constitutes. In case the central computer fails, the overall vehicle will fail. This must be already considered during design and development phase both in software and in hardware to satisfy the automotive industry's ASIL ratings [12].

There are many different technical approaches to enable the transformation to a zone-oriented E/E architecture and the resultant challenges. Those technical approaches will be presented in the next chapter and will be used to design methodical approaches for function abstraction and function centralization.

## 3      Technical Approaches for Centralization

Within this chapter, we will classify technical approaches to enable centralization from bottom to top of an ECU according to AUTOSAR's open and standardized software architecture. By this, one does not lose the relation to current architectures in vehicles. AUTOSAR's Classic platform representing established ECU architectures divides the ECU into its components from hardware towards application as can be seen in [13]. On the contrary, AUTOSAR's Adaptive platform already covers future vehicle ECU architectures based on for an example POSIX which is displayed in Fig. 3 [14]. The chapter will be closed by a cultural and a structural approach that can support automotive industry to realize the technical approaches and get closer to IT development and its toolchains to satisfy the needs of a software-driven vehicle architecture.

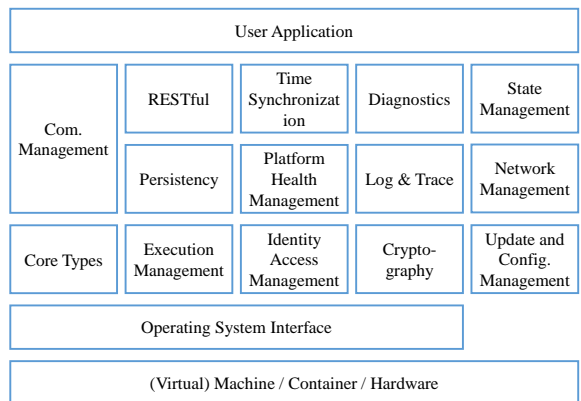

**Fig. 3.** R21-11 of AUTOSAR Adaptive following [14]

Centralized HPC ECUs need to consider a high amount of computing power to process the incidental data. Also the increase in demand over time of a software-driven vehicle must be considered due to further functionalities that may be flashed over the air (FOTA) during the vehicle's lifetime. To accommodate CPU-intensive tasks in a centralized ECU efficiently, hardware accelerators can reduce CPU load by overtaking specific tasks. A hardware accelerator is specialized hardware for dedicated tasks as it is for an example commonly used in graphics display of computers. Those can be switched on and off on demand and thus support the overall computing efficiency. [3]

As vehicle development moves forward with an increasing number of functions and new CPU-intensive technologies, single-core processing is no more sufficient to satisfy the requirements. A continuous increase in the CPU's frequency is restricted due to power dissipation limits [15]. Thus multi-core and manycore processors are seen as key enablers for centralization of E/E architectures. Those type of processors enable parallelism of task allocation and by this increase the available computing power. Multi-core can also support the temporal freedom of interference which is an important safety factor with regards to virtualization and usage of different operating systems (real time versus server client) on the same CPU. At the same time complexity will increase and backward compatibility for legacy platforms needs to be considered. For this purpose, there are various software mechanisms and different approaches to reach a balanced load of the cores and avoid misallocation while following computation sequences. Those will not be further discussed here and can be investigated in [4, 15, 16, 17].

Coming one step closer on the ECU level to the application, the microcontroller abstraction layer (MCAL) as lowest layer of the BSW comes into focus. It makes the upper software layers independent of the main processor and its on-chip peripherals like communication modules, memory, I/O and further. Operating systems interact with the hardware by the MCAL on an abstract level which allows programmers to code their SW device-independent. Thus, the MCAL increases SW portability to other ECUs which are closer to a potential central master ECU. [4]

ADAS and infotainment systems process and handle a high amount of data so that classic time-driven operating systems like AUTOSAR Classic in embedded ECU development reach its limits. The younger, POSIX-based AUTOSAR Adaptive standard enables automotive developers to implement in an event-driven, client-server oriented way as the adaptive platform enables dynamically linkage of services and clients during runtime. Those characteristics make it possible for ECUs to handle the complexity and computational load of recent and future vehicle functionalities. Nevertheless, AUTOSAR Adaptive will not displace AUTOSAR Classic which will still be applied for hard real-time, time-critical applications with low computational load [3, 18, 19]. Both may need to coexist within the same CPU for their related functions and applications bringing us to a further key enabler for centralization, the virtualization. Virtualization enables to run multiple operating systems on the same hardware and increases by this the potential to centralize two ECUs with different requirements as for an example deterministic and non-deterministic behavior on one common hardware [1].

The parallel execution of multiple operating systems on the same CPU requires the implementation of a hypervisor. The hypervisor is also called virtual machine monitor (VMM) and isolates the hardware from the virtual machines (VM) and its operating systems. The hypervisor monitors the hardware needs of the related VMs and manages the hardware distribution based on the needs. We distinguish two types of hypervisors: type 1 and type 2. The type 1 hypervisor, also called bare metal hypervisor, is the direct interface towards the hardware acting as host operating system. It manages the resources retrieval of the guest operating systems running on it. The type 2 hypervisor runs on the already available host operating system as an application requesting resources for the guest operating systems via host. The type 2 hypervisor performs pure software virtualization. Also virtualization requires same as for multi-core processing methods and patterns to ensure freedom of interference, memory protection, safety according to ISO 26262 [20] or communication regulation over VM borders. [1, 21]

Concluding with the impact of virtualization and operation systems on centralization of E/E architectures, middleware is reached on the way from ECU hardware to the application. It abstracts the application from the basis software, operating system and hardware. A middleware increases the flexibility of software components (SWCs) and enables that the software developer can fully focus on the function implementation without the need to consider lower layers. Smart middleware solutions will be the key enabler for the modern vehicle to access services of the IT domain and thus of the world wide web. By integration of a middleware, services can be properly translated and interpreted by both sides. While AUTOSAR Classic's time-driven middleware is called Runtime Environment, AUTOSAR Adaptive uses a middleware called ARA which handles communication between server and clients. [18, 19, 22]

In this context it gets clearer that smart middleware solutions are required to centralize and virtualize ECUs to ensure the connectivity of the software-driven vehicle.

This abstraction must also be supported by the application in form of SWCs that are designed flexible with defined interfaces. To reduce complexity and increase flexibility, the paradigm *divide and conquer* can support in designing applications out of decoupled microservices taking responsibility for smaller tasks. Microservices increase scalability and exchangeability of functions and software architectures. By orchestration, microservices can be reused for different applications. [2]

Putting a microservice together with everything it needs to run like for an example system libraries, configuration, tools and runtimes, a container is created representing an independent unit of deployment. Containers and containerization in general support realization of DevOps' practice continuous integration / continuous deployment (CI/CD). Containers can be seen as a unifying technical approach to design applications already in the beginning independent of the future hardware or host on that it will run. It gives the application a better potential for relocatability on a central master ECU and to manage increasing complexity when keeping the principle of microservices. [23]

While the presented technical approaches are already established in IT domain, automotive industry just started to introduce them with increasing digitalization in the automobile. Cultural and structural approaches can increase the potential to put all of the presented technical approaches into practice. Differences can be seen when comparing working methods and organizational structures of companies out of automotive industry with companies out of the IT domain. While in the automotive industry development according to the V model and vertical structures currently prevail, it is DevOps and horizontal structures in the IT domain.

DevOps, as the name indicates, combines development and operations by a number of various technical, methodical and cultural approaches. The goal is to fasten the development, the testing and the release process to improve software quality. By fusion of development and operations, feedback of operators and/or customers can directly feed into the development creating an endless loop which enables the principle continuous integration / continuous deployment (CI/CD). As the future vehicle will be more and more software-driven with the possibility to update single functions on demand, the automotive industry's working methods will evolve towards DevOps [2, 22, 24]. It is open if DevOps will replace the V model, if both will coexist or if a combination of both will be applied in future automotive development.

In addition the organizational structures, especially the communication structures, impact the developed system's architecture as Conway's Law states [3]. Applied to E/E architecture in vehicles, it means the E/E architecture represents the communication structure of the company. As current E/E architectures of vehicles are still characterized by vertical orientation, the vertical organizational structures could be one of the root causes according to Conway's Law. In other words, communication and organizational structures must be centralized to push forward centralization of E/E architectures. [3]

Taking the previously discussed technical approaches into account, a system and its functions must be in a first run separated into its single components before a methodical approach for function centralization can be set up. The next chapter defines a concept of function separation. Based on this, a methodical concept is developed which applies the technical solutions on abstracted functions with the aim to evaluate centralization potential. This methodical approach will support system designers and function engineers in decisions about how to distribute functions within a system with focus on the presented current trends and challenges in automotive development.

## 4 Development of Methodical Approaches

### 4.1 Methodical Approach for Function Abstraction

While SW can be designed in a flexible way and thus be location-independent as presented, HW has stronger dependencies due to its tasks to, for example, actuate charge inlet locks or sense charge inlet temperatures. It is not for each I/O possible to root analog signal lines over longer distances directly to a central master ECU which will be part of the validation and discussion part. Thus a classic decentralized function can serve as an example to abstract it into its elementary components.

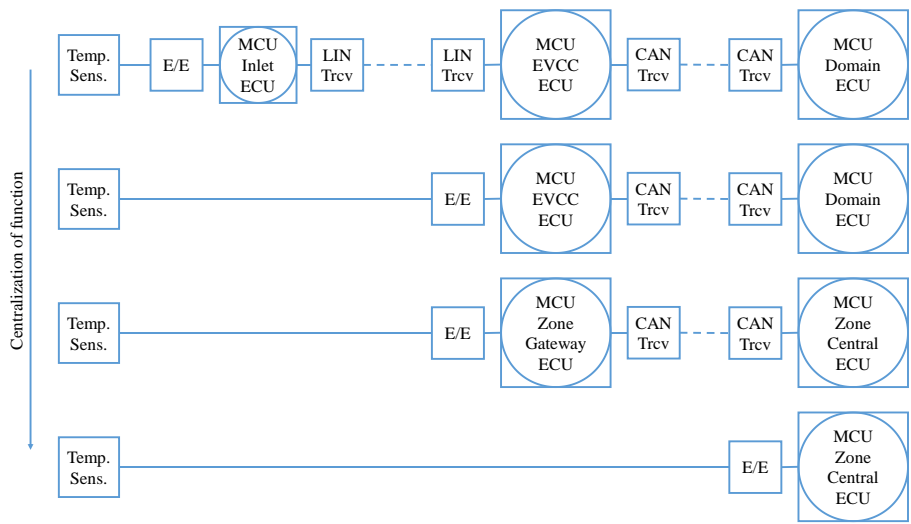

**Fig. 4.** Stepwise centralization of a derating function

Each ECU or controller project starts with HW as a basis. Necessarily some computing power and control logic is required by CPUs and Microcontroller Units (MCUs). Those will be the home for operating systems which distribute tasks from SWCs to the hardware kernels and cores or monitor and operate I/O electric circuits. The circuits will be also handled as the element HW which operates as interface between I/O like actuators or sensors and the processing unit serving the evaluation and control logic. In a next step I/O can be connected to the processing ECU in a centralized or decentralized way [12]. Thus the abstraction approach needs to consider direct analog I/O signaling lines (centralized) and digital communication protocols like LIN and CAN, Ethernet or even wireless according to the IEE 802.11 standards (decentralized). Centralized I/O has the advantage to reduce the number of ECUs as no gateways are needed and the information will be directly processed from a more powerful and bigger centralized master ECU [12]. Nevertheless, it also implies disadvantages that will be listed in the discussion.

By this the four chosen elementary components HW, SW, I/O signaling line and I/O COM are already sufficient to abstract functions which enables to evaluate the function's potential for centralization. Fig. 4 shows the generic illustration of the elementary components which will be used for the validation based on a temperature current derating function which is used in modern hybrid and electric vehicles.

### 4.2 Methodical Approach for Function Centralization

The methodical approach starts from a domain-oriented E/E architecture which is handled as state of the art. By centralizing as much as logic as possible in the domain controller, the domain controller evolves towards a central master computer as in best case only smart actuators and sensors will be left. In a next elaboration, a methodical concept can be developed to centralize the logic of the domains into cross-domains or into one central computer so that the domain controllers remain as simple translating zone gateway controllers.

Fig. 5 shows the UML activity diagram which we created based on the technical solutions and the basic idea of centralization to be capable of manage current trends and challenges. The aim is to move forward in function centralization from a decentralized and distributed domain-oriented function architecture. The evaluation must be performed for each of the domain's function to accommodate as much as functionality within the domain ECU itself evolving to a central master computer.

### 4.3 Validation of Methodical Approach

To validate the methodical approach shown in the UML activity diagram, a temperature current derating function which is typical for hybrid and electric vehicles is used. Based on temperatures of sensors within the charging inlet, the charging current drawn by the vehicle will be regulated to prevent damage and injury. This function has an ASIL rating of B according to ISO 26262 [20]. The relevance of this information on the potential of centralization will be part of the discussion.

In a first step, the current function architecture must be abstracted according to the presented methodical approach for function abstraction in chapter 4.1. Fig. 4 shows on top the actual function architecture with its elementary hardware components, the corresponding circuits and the ECUs and MCUs which run the SWCs. Top down the function gets centralized until first its architecture equals the principle of a zone-oriented E/E architecture as described in chapter 2.3 and second its architecture is centralized as much as possible. The technical approaches out of chapter 3 and the UML activity diagram out of chapter 4.2 are used therefore.

The main logic consists of two parts. Part one evaluates the temperature sensors by an electric circuit and provides the temperature values of the AC and the DC pins. This evaluation is mandatory as according to IEC 62196, it must be ensured that the maximum temperature at touchable, non-metallic parts shall not exceed 85 °C. Part two of the main logic processes the temperature values and calculates the potential current which can be drawn. The derating function can be seen as a quality performance function. The charging can be also performed with a continuous, maximum current. By this the threshold of 85°C might be reached quickly and charging needs to be stopped even if the target state of charge of the vehicle is not reached yet. Thus the derating function shall regulate the current so that the charging process does not directly run in the threshold and needs to be stopped. Instead the charging shall be continued with a lower current to reach the optimal time span until the target state of charge is reached.

Part one of the main logic cannot be shifted freely due to its connection and reliance on the sensors and corresponding electric circuit. Whereas part two of the main logic can be designed as a kind of microservice according to chapter 3. It has clear inputs like the temperature values and clear outputs meaning the actual maximum current based on its calculation. This fact makes the SWC or microservice *temperature current derating* flexible and independent of its location.

As can be seen on top of Fig. 4, the function *temperature current derating* is allocated to three different ECUs. Following the UML activity diagram, part two of the main logic above can be shifted in an ECU closer to the domain controller. In the first step this will be the ECU *EVCC*. The I/O of the sensors including their electric circuit is centralized by a direct analog signaling line from sensor towards *EVCC*. By this, the function itself does no more have the need for the ECU *Inlet*. It does only apply for this dedicated function so that the ECU *Inlet* cannot be removed in general. This is also the reason why the UML activity diagram must be executed for every function $x$ of a domain. The function with the least potential for centralization limits the removal of components. The new architecture of the function is the second from the top of Fig. 4.

Going one step further with regards to a zone-oriented E/E architecture, the ECU *EVCC* will be converted into a simple zone gateway ECU by shifting part 2 of the main logic into the domain ECU which shall evolve to a central computer. The zone gateway ECU serves now as a simple smart sensor which monitors and translates for the domain controller. Below the zone gateway ECU, further sensors and actuators could be placed which will be monitored and processed for the domain controller. The new function architecture is displayed in the third row of Fig. 4. In the last step of the UML diagram, even the zone gateway ECU will be removed by centralizing once more the I/O directly to its domain controller. This leads to a shift of the electric circuit into the domain controller reducing also the number of communication lines. This most centralized function architecture is shown in the last row of Fig. 4. As a further centralization of the function *temperature current derating* is no more possible, a discussion on the advantages, disadvantages and challenges of the different variants will follow.

## 5    Discussion

The most obvious advantage of the fourth and most centralized function design is the reduction of components, electric circuits and communication channels. Development costs can be saved, busload and latency decrease, reliability increases and the scalability is less restricted in a software-driven central computer. [12]

Applying the technical approaches of chapter 3 on the central computer, functions can be handled as known from smartphones. New functions can be added and outdated functions can be updated continuously as it is done in IT domain principles like for an example CI/CD. [22]

At the same time, the centralization reduces modularity and flexibility as a clear advantage of actual decentralized architectures. It impacts especially the HW design of the domain or central controller which by this may include components that are only needed in certain vehicle variants. In addition to that, centralized I/O also increases the wire lengths and number of cut leads (point-to-point wire connections) as sensors and actuators cannot be connected as close as possible to the nearest bus. [12]

The longer the wire, the higher the impact of environmental influences and signal losses. In a vehicle, especially environmental impacts and EMC can result in loss of signal accuracy over longer distances. With a view to proper signal qualities and reliabilities to satisfy ASIL ratings and the ISO 26262 [20], twist and shield might be necessary which increases material, evaluation and certification costs. To eliminate those disadvantages, the zone-oriented E/E architecture with decentralized I/O connected by communication busses can be chosen. This function design with a zone gateway ECU equals the third design in Fig. 4.

Another risk of a fully centralized E/E architecture design is the single point of failure. The central computer in a vehicle needs to fulfill ASIL ratings up to ASIL D to avoid damage and injury. By this the central computer must show redundancy for power supply, communication and further. [20] It is a high challenge also on SW level which requires appropriate counter measures to ensure the timely achieving of the vehicle's or system's safe state.

## 6   Conclusion

Within this paper, we propose a methodical approach supporting system and function engineers in centralization of E/E architectures. Actual technical solutions out of research were collected, investigated and merged into an UML activity diagram by means of which different function designs can be worked out and centralized. Nevertheless, centralization cannot be seen as general panacea for each function of an E/E architecture. In general, it will improve manageability and reduce complexity of future technologies like ADAS or infotainment systems which are characterized by a high amount of code lines and data. Those might not be manageable otherwise in the near future. However, there are also disadvantages when centralizing functions which must be elaborated in detail especially when it refers to conventional functions and functions that do not require high computing power or high amounts of data.

In future work, a methodical approach that shifts the functions of the domain controllers into a central master computer and transforms domain controllers into zone gateway controllers can be worked out. Additionally, the future of the V model under the influence of DevOps needs to be revised. By designing integration concepts of DevOps into the V model, the impact on the flexibility and quality of the development process can be evaluated.

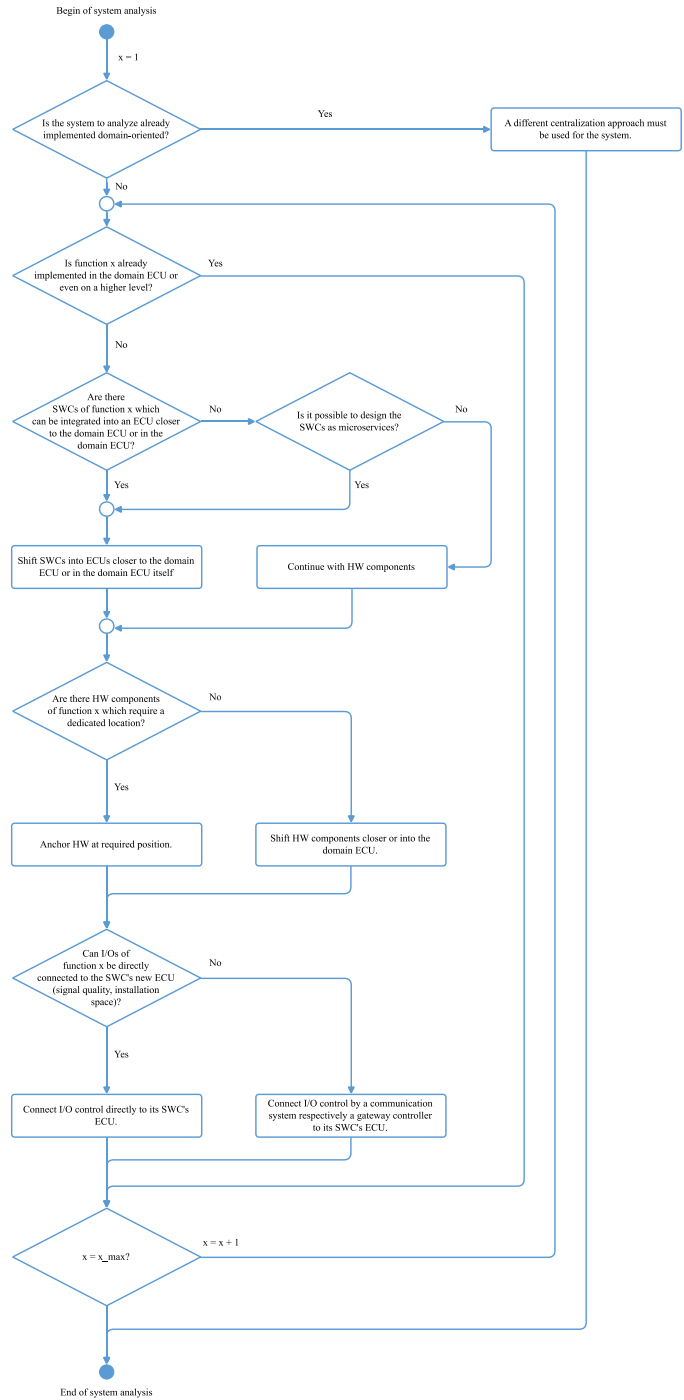

**Fig. 5.** UML activity diagram to centralize distributed, domain-oriented function designs